\documentclass[twocolumn,groupedaddress,aps,pra,10pt,floatfix]{revtex4-2}
\usepackage[colorlinks,linkcolor=blue,anchorcolor=blue,citecolor=blue,bookmarksnumbered]{hyperref}
\usepackage{graphicx,epsfig,float,epstopdf}
\usepackage{amssymb,amsmath,subeqnarray,mathrsfs,amsfonts,bm,makecell,cases,color,extpfeil,lipsum}
\usepackage[normalem]{ulem}

\allowdisplaybreaks

\begin{document}

\title{Enhanced spontaneous-emission and localization in a non-Hermitian Lieb lattice}

\author{Lu Qin$^{1}$}
\author{Shengqiang Chen$^{1}$}
\author{Yingying Zhang$^{1}$}
\author{Zeyun Shi$^{2*}$}
\author{Zunlue Zhu$^{1}$}
\author{Chao Hang$^{3,4*}$}
\author{Xingdong Zhao$^{1}$}

\thanks{shizy@huat.edu.cn\\
chang@phy.ecnu.edu.cn\\
phyzhxd@gmail.com}

% \author{Lu Qin$^{1}$\orcidlink{0000-0001-6977-2680}}
% \author{Hairu Zhai$^{1}$}
% \author{Zeyun Shi$^{2}$\textsuperscript{*}\orcidlink{0000-0003-4266-0691}}
% \author{Yingying Zhang$^{1}$\textsuperscript{$\dagger$}}
% \author{Zunlue Zhu$^{1}$}
% \author{Wumimg Liu$^{1}$}
% \author{Xingdong Zhao$^{1}$\textsuperscript{$\ddagger$}\orcidlink{0000-0001-9533-7279}}

\affiliation{
$^1$School of Physics, Henan Normal University, Xinxiang 453007, China\\
$^2$School of Artificial Intelligence, Hubei University of Automotive Technology, Shiyan 442000, China\\
$^3$Institute of Quantum Science and Precision Measurement, East China Normal University, Shanghai 200241, China\\
$^4$Collaborative Innovation Center of Extreme Optics, Shanxi University, Taiyuan, Shanxi 030006, China}

\date{\today}

\begin{abstract}

We propose a physical scheme for the realization of a non-Hermitian Lieb lattice in an optically pumped four-level $N$-type cold atom medium. The atomic medium provides a complex susceptibility that can be used to engineer the dispersive and dissipative on-site parameters of the Lieb lattice. We investigate a third-order exceptional point (EP3) and a second-order exceptional line (EL2) through numerical calculations and simulations. 
We used the Petermann factor (PF) to verify the coalescence and non-orthogonality of the eigenmodes at  EP2 and EP3.
% We also examine the Petermann factors at the EP2 and EP3. 
In addition, we study the spontaneous-emission at EPs, in connection with the local density of states (LDOS). We find that spontaneous-emission can be enhanced by factors of 4 and 8 at the EP2 and EP3, respectively, in passive system. We further investigate localization in non-Hermitian lattices, which can be realized at arbitrary lattice sites by suitably tuning the loss of the lattice sites. Our results are useful for applications requiring enhanced and controlled emission, such as quantum sensing and imaging.

\end{abstract}

\maketitle

%% 一、引言 %%%%%%%%%%%%%%%%%%%%%%%%%%%%%%%%%%%%%%%%%%%%%%%%%%%%%%%%%%%%%%%
\section{Introduction}\label{sec:level1}

Open systems with gain, loss, or dissipative coupling are naturally described by non-Hermitian operators and can exhibit spectral and dynamical behavior unavailable in closed Hermitian systems~\cite{NH03,NH07,NH09,ADD03_RotterBird2015,ADD04_ElGanainy2019}. Among their characteristic degeneracies are exceptional points (EPs), where eigenvalues and their eigenvectors coalesce~\cite{NH04,ADD01_Heiss2004,ADD02_Berry2004}. This eigenvector coalescence produces self-orthogonality and strong modal nonorthogonality, commonly quantified by the Petermann factor (PF), and underlies a range of effects in wave transport, sensing, and emission control~\cite{NH22,NH30,NH69,ADD30_Berry2003,ADD32_Lee2009,NH24,NH28,Hodaei2017EnhancedSensitivity,Zhang2025Tunable}.

Many studies have focused on second-order EPs (EP2s), where two eigenvalues and two eigenmodes coalesce. Higher-order EPs involve the coalescence of more eigenmodes and are associated with higher-order poles in response functions, giving rise to richer spectral line shapes and modal dynamics~\cite{NH18,NH19,NH20,ADD05_HeissWunner2016,ADD07_Zhong2018,ADD09_Zhong2020,ADD10_Xiao2019}. A third-order EP (EP3), in particular, requires three eigenvalues and all three corresponding eigenvectors to coalesce. Its realization therefore demands more than observing a crossing of complex eigenvalues: the degeneracy and  dimension of its eigenspace must also be established~\cite{ADD05_HeissWunner2016,ADD08_Laha2020,ADD11_KulligWiersig2019,ADD13_Kullig2018,ADD14_Zhang2023,NH15,ADD12_Wiersig2022}.

A Lieb lattice is a natural setting for this problem. Its unit cell contains three iequivalent sites connected by nearest-neighbor couplings, forming a three-band system in which all bands can participate in a common degeneracy~\cite{NH32,NH33,NH34,ADD18_Taie2015,ADD19_Slot2017,ADD20_Apaja2010}. The three-site basis is the minimal lattice structure capable of hosting an EP3, while the on-site terms dependent on sublattice allow the control of gain and loss without changing the coupling. This makes the Lieb lattice a useful platform for examining both isolated higher-order degeneracies and second-order exceptional lines (EL2s).

A practical route to implementing such complex Lieb lattice  is provided by electromagnetically induced transparency (EIT) in a cold atomic medium. The optical response of an EIT medium can be tuned using laser detunings and Rabi frequencies, and spatially patterned fields can make that response sublattice dependent~\cite{NH51,NH53,NH54,ADD23_Boller1991,Sheng2013,Hang2013,Qin2023OE,Qin2019ControllablePT,Qin2027FOP}. In a four-level optically pumped $N$-type configuration, the linear susceptibility is complex: its real part controls the dispersive refractive-index shift, whereas its imaginary part determines absorption or gain~\cite{ADD29_Kang2003,Sheng2013,Qin2019ControllablePT}. Patterning the control laser field thus offers an all-optical means of mapping the atomic response onto the on-site parameters of a non-Hermitian Lieb lattice~\cite{Feng2023LossDifference}.
Because the gain and loss of the Lieb lattice are tunable via the atomic system parameters, the system's gain and loss can be adjusted by varying quantities such as the laser field detuning~\cite{Sheng2013,Qin2019ControllablePT}. Consequently, site-dependent localization can be achieved by adjusting the loss or gain at individual lattice sites~\cite{Feng2023LossDifference}.

% Because atomic system parameters are readily tunable, the gain and loss of the system can be tuned by changing the field detuning, among other knobs. Consequently, site-dependent localization can be achieved by tuning the loss or gain at individual lattice sites~\cite{Feng2023LossDifference}.

The spectral consequences of mode coalescence also motivate a careful distinction between modal nonorthogonality and the response experienced by a local spontaneous-emission. A divergent PF diagnoses the vanishing left--right modal overlap at an EP, but it is not itself an emission rate. The spontaneous-emission is determined by the electromagnetic response projected at the emitter and is related to the local density of states (LDOS), which can be evaluated from the appropriate local Green's function element~\cite{Pick2017GeneralTheory,NH65,NH66,ADD34_Kristensen2020}. At an EP, the ordinary eigenmode expansion is incomplete; a Jordan-chain representation is required and introduces higher-order poles into the Green's function~\cite{NH64,NH70,Ferrier2022UnveilingEmission}. The resulting LDOS depends on the local channel, so its peak enhancement need not equal a universal enhancement of spontaneous-emission for an arbitrary dipole.

%% fig1 %%%%%%%%%%%%%%%%%%%%%%%%%%%%%%%%%%%%%%%%%%%%%%%%%%%%%%%%%%%%%%%
\begin{figure*}[ht]
\includegraphics[width=0.85\linewidth]{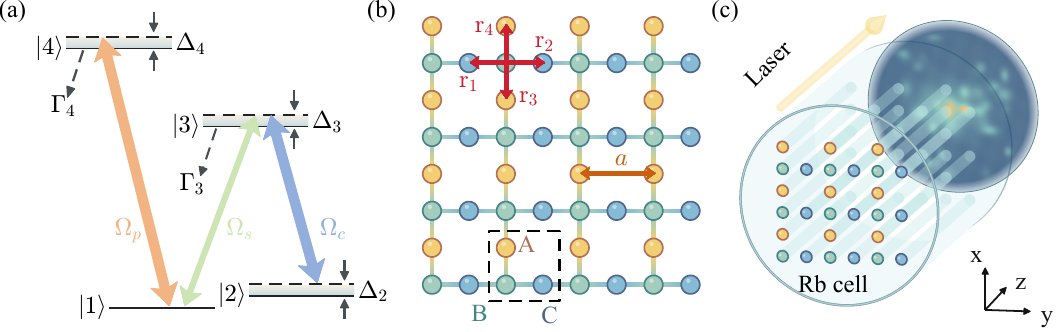}
\caption{
Physical model. (a) Energy level diagram and excitation scheme of the
$N$-type four-level atomic gas. $|\alpha\rangle$ ($\alpha=1,\,2,\,3,\,4$) denote the atomic
states, and $\Delta_{\alpha}$ are the corresponding detunings. $\Gamma_3$ and
$\Gamma_4$ denote the spontaneous decay rates of $|3\rangle$ and $|4\rangle$,
respectively. $\Omega_s$, $\Omega_c$, and $\Omega_p$ are the half Rabi frequencies
of the weak signal field (green), the strong control field (blue), and the strong
pump field (orange), respectively. (b) Two-dimensional (2D) Lieb lattice. The dashed black box marks a primitive cell
containing the three sublattice sites $A$ (yellow), $B$ (mint green), and $C$
(dark blue). The lattice constant is $a$, and $\mathbf{r}_{1,\,2,\,3,\,4}$ denote the
coupling directions in the $x$-$y$ plane. (c) Experimental scheme. A Lieb-lattice refractive-index distribution is induced inside the vapor cell by the spatial profile of the control laser field, whose two-photon detuning $\Delta_2$ varies from site to site. All laser fields propagate along the $z$ axis.
}
\label{fig1}
\end{figure*}
%% fig1 %%%%%%%%%%%%%%%%%%%%%%%%%%%%%%%%%%%%%%%%%%%%%%%%%%%%%%%%%%%%%%%

In this work, we propose a scheme for realizing a non-Hermitian Lieb lattice in an optically pumped four-level cold atomic medium and analyze its three-band tight-binding model. We derive the conditions for an EP3 and identify EL2s and EP3s in the band structure. 
% For the purely dissipative spectral configuration, the $A$ and $B$ sublattices are neutral and the $C$ sublattice is lossy. 
We characterize the associated mode coalescence using PF, then construct Jordan chains to obtain the Green's functions and local spectral responses at EP2 and EP3. For the local channels and parameters studied here, the calculated LDOS peak is 4 times (EP2) and 8 times (EP3) the peak of a single nondegenerate resonance. 
% These are model- and channel-specific LDOS comparison in passive system; predicting the rate of a particular emitter additionally requires its position and coupling to the lattice channels.
 Consequently, spontaneous-emission can be enhanced in the Lieb lattice system that we propose.
We also investigated light propagation with independently tunable sublattice loss parameters. By varying the on-site loss parameters while launching a localized excitation, we show how the relative field intensity in a central unit cell can be redistributed among its three sites. 
% Because a localized input excites a range of Bloch wave vectors, this propagation study characterizes the lattice's broader real-space response rather than the dynamics of a single EP. 
% Together, 
The spectral and propagation analyses illustrate how an EIT medium can provide a tunable platform for exploring higher-order EP physics and controlling local optical responses in a non-Hermitian Lieb lattice.

The paper is arranged as follows. In Sec.~\ref{model}, we introduce a physical model for realizing a non-Hermitian Lieb lattice. In Sec.~\ref{sec:complex_band}, we discuss the EPs and the ELs, and demonstrate the divergence of the PF at an EP. In Sec.~\ref{sec:LDOS_enhancement}, we investigate the enhancement of the local density of states at an EP. In Sec.~\ref{sec:propagation_dynamics}, we show how tuning the loss distribution enables localization and energy exchange. Finally, we summarize our findings in Sec.~\ref{conclusion}.

\section{Physical model}\label{model}
\subsection{Four-level $N$-type cold atom system}

We consider a cold, dilute $N$-type four-level $^{85}$Rb atomic gas, which  consists of two hyperfine ground states $|1\rangle=|5^2S_{1/2},F=2\rangle$, $|2\rangle=|5^2S_{1/2},F=3\rangle$, and two excited states
$|3\rangle=|5^2P_{1/2}\rangle$ and $|4\rangle=|5^2P_{3/2}\rangle$~\cite{Steck,Feng2023LossDifference}. A weak signal field $\mathbf{E}_s$,
% of center frequency
% $\omega_s$ and wave number $k_s=\omega_s/c$, 
a strong control field
$\mathbf{E}_c$,
% of frequency $\omega_c$ and wave number $k_c=\omega_c/c$,
and a strong pump field $\mathbf{E}_p$
% of frequency $\omega_p$ and wave
% number $k_p=\omega_p/c$ 
couple the transitions
$|1\rangle\leftrightarrow|3\rangle$,
$|2\rangle\leftrightarrow|3\rangle$, and
$|1\rangle\leftrightarrow|4\rangle$, respectively; see Fig.~\ref{fig1}(a).
The electric-field vector can be written as
$\mathbf{E} = \mathbf{E}_s + \mathbf{E}_c + \mathbf{E}_p
= \sum_{l=s,\,c,\,p} \mathbf{e}_l \mathcal{E}_l \exp[i(k_l z - \omega_l t)] + \text{c.c.}$,
where $\mathbf{e}_l$, $\mathcal{E}_l$, $\omega_l$, $k_l$ are the polarization unit vector, envelope, frequency, and wave
number of the laser field, respectively. 
% $\omega_l$ and wave number $k_l=\omega_l/c$,
All laser fields propagate along
$z$ to minimize Doppler effect.
The scheme in Fig.~\ref{fig1}(a) is an optically pumped EIT system: a standard three-level $\Lambda$ configuration formed by $|1\rangle$, $|2\rangle$, and $|3\rangle$ and coupled by the signal and control fields is supplemented by the optical-pumping transition $|1\rangle\leftrightarrow|4\rangle$, which provides Raman gain for the signal field~\cite{NH53,ADD29_Kang2003}.

Under the electric-dipole and rotating-wave approximations, the Hamiltonian in interaction picture is
$\hat{H}_{\rm int} = -\hbar \sum_{\alpha=1}^4 \Delta_\alpha
|\alpha\rangle\langle \alpha| - \hbar[\Omega_p|4\rangle\langle 1| +
\Omega_s|3\rangle\langle 1| + \Omega_c|3\rangle\langle 2| + \text{H.c.}]$,
where $\Delta_1=0$, $\Delta_2=\omega_s-\omega_c-(E_2-E_1)/\hbar$ is the
two-photon detuning, and $\Delta_3=\omega_c-(E_3-E_1)/\hbar$ and
$\Delta_4=\omega_p-(E_4-E_1)/\hbar$ are the one-photon detunings. Here $E_\alpha$ is the eigenenergy
of state $|\alpha\rangle$; $\Omega_s = (\mathbf{e}_s \cdot \mathbf{p}_{13})\mathcal{E}_s/(2\hbar)$,
$\Omega_c = (\mathbf{e}_c \cdot \mathbf{p}_{23})\mathcal{E}_c/(2\hbar)$, and
$\Omega_p = (\mathbf{e}_p \cdot \mathbf{p}_{14})\mathcal{E}_p/(2\hbar)$ are the
half Rabi frequencies of the signal, control, and pump fields, respectively,
with $\mathbf{p}_{\alpha\beta}$ the electric-dipole matrix element associated
with the transition $|\alpha\rangle \leftrightarrow |\beta\rangle$.

The dynamics of the atoms is governed by the optical Bloch equation
\begin{equation} \label{eq:Bloch}
\frac{\partial \rho}{\partial t} = -\frac{i}{\hbar}[\hat{H}_{\rm int},\rho]
- \Gamma[\rho],
\end{equation}
where $\rho$ is the $4\times4$ density matrix (with elements $\rho_{\alpha\beta}$;
$\alpha,\,\beta=1-4$) describing the atomic populations and
coherences, and $\Gamma$ is the relaxation matrix accounting for spontaneous
emission and dephasing.
The explicit form of Eq.~(\ref{eq:Bloch}) is given in
Appendix~\ref{app:bloch_solution}.

The signal field susceptibility is
$\chi_s={\mathcal{N}_a|\mathbf{e}_s\cdot\mathbf{p}_{13}|^2\rho_{31}}$ $/{(\varepsilon_0\hbar\Omega_s)}$
~\cite{Sheng2013,Qin2019ControllablePT}.
% \begin{equation}
% \chi_s=\frac{\mathcal{N}_a|\mathbf{e}_s\cdot\mathbf{p}_{13}|^2\rho_{31}}{\varepsilon_0\hbar\Omega_s}.
% \label{eq:linear_susceptibility}
% \end{equation}
For a dilute vapor, $|\chi_s|\ll1$, so $n=\sqrt{1+\chi_s}\approx1+\chi_s/2$. Writing $\chi_s=\chi_s'+i\chi_s''$ and $n=n_0+n_R+i n_I$, with the vacuum background $n_0=1$, gives $n_R\approx\chi_s'/2$ and $n_I\approx\chi_s''/2$.
Because the signal is weak, we treat $\Omega_s$ perturbatively.
We focus on the steady-state response, neglecting the time derivatives in Eq.~\eqref{eq:Bloch}. This approximation is appropriate when the signal duration is much longer than the atomic relaxation time.  The steady-state solution is given in Appendix~\ref{APP:solution}.

% A key advantage of this non-Hermitian refractive index is its in situ tunability,
% which arises from the multi-level structure of the atomic system

A key advantage of the multi-level atomic system is its ability to realize a non-Hermitian refractive index with flexible in situ tunability
~\cite{Hang2013,Qin2019ControllablePT,Zhang2016,Sheng2013,Qin2023OE}.
Moreover, the refractive index is closely related to optical susceptibility.
% {\color{red}
% One possible realization uses 
% % $^{85}$Rb atoms, with
% % $|1\rangle=|5^2S_{1/2},F=2\rangle$, $|2\rangle=|5^2S_{1/2},F=3\rangle$,
% % $|3\rangle=|5^2P_{1/2}\rangle$, and $|4\rangle=|5^2P_{3/2}\rangle$.
% $^{87}$Rb atoms, with
% $|1\rangle=|5^2S_{1/2},F=1\rangle$, $|2\rangle=|5^2S_{1/2},F=2\rangle$,
% $|3\rangle=|5^2P_{1/2}\rangle$, and $|4\rangle=|5^2P_{3/2}\rangle$.
% % The spontaneous emission rates are
% % $\Gamma_1=\Gamma_2\sim0$ (the $|1\rangle\leftrightarrow|2\rangle$ transition
% % being dipole-forbidden within the ground-state hyperfine manifold),
% % $\Gamma_{3}=2\pi\times 5.75$ MHz, and $\Gamma_{4}=2\pi\times6.07$ MHz~\cite{steck}.
% }
%%%%%
% Other system parameters $\Delta_3=\Delta_4= 0$,  $\Gamma_{3}=2\pi\times 5.75$ MHz, $\Gamma_{4}=2\pi\times6.07$ MHz, $\Omega_s=2\pi\times0.37$ MHz, $\Omega_p=2\pi\times 4$ MHz, $\Omega_{c0}=2\pi\times2$ MHz, and
% $\mathcal{N}_a=1.8\times 10^{13}~{\rm cm}^{-3}$.
% Because $|1\rangle$ and $|2\rangle$ are two hyperfine levels of the same ground-state, the
% electric-dipole transition between them is forbidden, and the spontaneous
% population decay rate of $|2\rangle$ can be taken as
% $\Gamma_2\simeq0$.
%%%%

Figure~\ref{fig2}(a) shows the real and imaginary parts of $\chi_s$ as functions of the two-photon detuning $\Delta_2$. The neutral point $N$ at $\Delta_2=-29.23~\mathrm{MHz}$ lies close to the zero-absorption condition $\operatorname{Im}[\chi_s]=0$ and therefore gives an approximately neutral optical response. $\operatorname{Im}[\chi_s]>0$ corresponds to loss, whereas $\operatorname{Im}[\chi_s]<0$ corresponds to gain. Figure~\ref{fig2}(b) shows the loss (yellow) region to the left of $N$ and the gain (blue) region to its right.
Assigning different values of $\Delta_2$ to the $A$, $B$, and $C$ sublattices [Fig.~\ref{fig1}(b)] therefore enables site-by-site control of the complex susceptibility and the associated gain, neutral, or loss response. This tunability provides the basis for realizing the effective non-Hermitian potential in the atomic vapor cell. 
% The orange triangle and hexagram mark the lossy  and neutral operating point at \(\Delta_2=-37\,\mathrm{MHz}\) and \(-29.23\,\mathrm{MHz}\). 

\subsection{Effective Non-Hermitian Lieb Lattice}

% We map the atomic medium onto a photonic lattice described by a tight-binding model. Particularly, this can be realized by applying another far-off resonance laser field, which is used to trap the atoms in each point of the Lieb lattice.
% Figure~\ref{fig1}(b) shows the site-selective Lieb lattice: the dashed black box marks a primitive unit cell containing the three sublattices $A$, $B$, and $C$, and $\mathbf{r}_{1,2,3,4}$ indicate the nearest-neighbor coupling directions. 
% Figure~\ref{fig1}(c) illustrates a possible implementation in a rubidium vapor cell, where a spatially patterned control field with different two-photon detuning $\Delta_2$ creates the Lieb-lattice refractive index profile. 
% Engineering the complex optical response at each site allows the sublattices to be tuned between gain, loss,  and neutral; see the Fig.~\ref{fig2}(b).

Figure~\ref{fig1}(b) shows the site-selective Lieb lattice. The dashed black box encloses a primitive unit cell containing the three sublattices $A$, $B$, and $C$, and $\mathbf{r}_{1,2,3,4}$ denote the nearest-neighbor displacement vectors. As illustrated in Fig.~\ref{fig1}(c), the lattice can be optically induced in a rubidium vapor cell by a control field with a spatially patterned transverse intensity. In EIT, the control field modulates the linear susceptibility experienced by the weak probe field. The intensity pattern and the two-photon detuning $\Delta_2$ can be chosen so that  the probe refractive index forms an array of guiding channels arranged in the Lieb geometry.

% Engineering the complex optical response at each site allows the sublattices to be tuned between gain, loss,  and neutral; see the Fig.~\ref{fig2}(b).

\begin{figure}[ht]
  \includegraphics[width=0.9\linewidth]{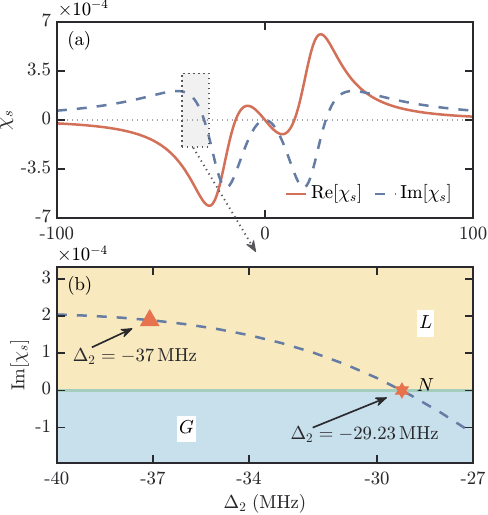}
  \caption{
  (a) Real (red solid line) and imaginary (blue dashed line) parts of the optical susceptibility \(\chi_s\) as functions of the two-photon detuning \(\Delta_2\). (b) 
  % Imaginary part of the optical susceptibility \(\chi_s\) as function of  \(\Delta_2\). 
  Enlarged view of the imaginary part within the dashed box in (a). The yellow and blue regions correspond to loss (\(\operatorname{Im}[\chi_s]>0\)) and gain (\(\operatorname{Im}[\chi_s]<0\)), respectively. 
  The orange triangle and hexagram mark the loss  and neutral operating point at \(\Delta_2=-37\,\mathrm{MHz}\) and
  \(-29.23\,\mathrm{MHz}\), respectively. 
  % The orange triangle marks the lossy operating point at \(\Delta_2=-37\,\mathrm{MHz}\), while the orange hexagram \(N\) marks the approximately neutral operating point at \(\Delta_2=-29.23\,\mathrm{MHz}\). 
  % These two detunings are used to establish the lossy response of the \(C\) sublattice and the neutral responses of the \(A\) and \(B\) sublattices, respectively.
  }
  \label{fig2}
  \end{figure}

Sublattice-selective control of the optical response can be implemented by shifting the effective two-photon detuning, for example, through an AC Stark shift induced by a spatially patterned, far-detuned auxiliary field. The imaginary part of the effective refractive index determines whether the probe experiences gain, loss, or neither at a given site, as illustrated in Fig.~\ref{fig2}(b). 
% A site with net gain requires the atomic level scheme and pumping conditions to support probe amplification.

In the weak probe and paraxial regimes, the transverse probe field can be expanded in localized channel modes. We retain one dominant mode per site, assuming that the guiding contrast is sufficient to confine the modes and that intersite coupling and site-selective modulation weakly mix the retained modes with other transverse modes. The evanescent overlap between neighboring channels gives rise to the $A$--$B$ coupling along $y$ and the $B$--$C$ coupling along $x$. We neglect farther-neighbor couplings and take the two nearest-neighbor coupling strengths to be approximately equal and predominantly real. This reduction applies when the site-selective modulation acts mainly on the on-site response and has a small effect on the mode overlaps. Within this tight-binding approximation, the effective Bloch Hamiltonian of the Lieb lattice reads~\cite{Zhang2016DispersionLieb,ADD15_Xiao2020}
% Within the tight-binding approximation, the effective Hamiltonian of the Lieb
% lattice in momentum space reads~\cite{Zhang2016DispersionLieb,ADD15_Xiao2020}
\begin{align}
\mathcal{H}
=
\begin{pmatrix}
n_A & g_y & 0 \\[2pt]
g_y & n_B & g_x \\[2pt]
0 & g_x & n_C
\end{pmatrix},
\label{eq:LiebHamiltonian}
 \end{align}
with the momentum-dependent coupling functions
\begin{equation}
\begin{aligned}
g_x
&=
\kappa\sum_{\nu=1}^{2} e^{i\mathbf{k}\cdot\mathbf{r}_{\nu}}
=
2\kappa\cos\left(\frac{k_xa}{2}\right),\\
g_y
&=
\kappa\sum_{\nu=3}^{4} e^{i\mathbf{k}\cdot\mathbf{r}_{\nu}}
=
2\kappa\cos\left(\frac{k_ya}{2}\right).\notag
% \label{eq:momentum_couplings}
\end{aligned}
\end{equation}
Here, $\kappa$ is the nearest-neighbor coupling strength, and
$\mathbf{k}=(k_x,k_y)$ is the two-dimensional wave vector. The displacement
vectors are $\mathbf{r}_1=(-a/2,0)$, $\mathbf{r}_2=(a/2,0)$,
$\mathbf{r}_3=(0,-a/2)$, and $\mathbf{r}_4=(0,a/2)$. Thus $g_x$
couples the $B$ and $C$ sublattices along $x$, while $g_y$ couples the $A$
and $B$ sublattices along $y$.
The diagonal entries $n_A$, $n_B$, and $n_C$ are the effective refractive indices at the corresponding sites and act as on-site potentials. 
% In our convention, an on-site term $n_0+n_R-i\gamma$ with $\gamma>0$ represents optical loss (amplitude decay), whereas $n_0+n_R+i\gamma$ represents gain.

%% fig2 %%%%%%%%%%%%%%%%%%%%%%%%%%%%%%%%%%%%%%%%%%%%%%%%%%%%%%%%%%%%%
%% fig2 %%%%%%%%%%%%%%%%%%%%%%%%%%%%%%%%%%%%%%%%%%%%%%%%%%%%%%%%%%%%%

As shown in Fig.~\ref{fig2}, the real part of the susceptibility Re[$\chi_s$] is small compared with the vacuum background $n_0$, so the real refractive index is approximately unity on each site. The sublattice contrast is therefore determined primarily by the imaginary part, $n_I=\operatorname{Im}[\chi_s]/2$, which is controlled by $\Delta_2$. In our analytical treatment, we consider a purely dissipative configuration (passive system), $n_A=n_B=n_0$ and 
$n_C=n_0-i\gamma$, with $\gamma=0.195$; 
thus, the $A$ and $B$ sublattices are neutral, while the $C$ sublattice is loss. The corresponding two-photon detunings are $\Delta_{2(A)}=\Delta_{2(B)}=-29.23~\mathrm{MHz}$ and $\Delta_{2(C)}=-37~\mathrm{MHz}$. 
They are marked by the orange triangle and hexagram in Fig.~\ref{fig2}(b).
Other parameters are $\Delta_3=\Delta_4=0$, $\Omega_p=4\times2\pi~\mathrm{MHz}$, $\Omega_c=1.5\times2\pi~\mathrm{MHz}$, $\Omega_s=0.1\times2\pi~\mathrm{MHz}$, $\mathbf{p}_{13}=2.54\times10^{-27}~\mathrm{C\cdot cm}$, and $\mathcal{N}_a=2\times10^{11}~\mathrm{cm}^{-3}$.
The spontaneous emission rates are
$\Gamma_1=\Gamma_2\sim0$ (the $|1\rangle\leftrightarrow|2\rangle$ transition
being dipole-forbidden within the ground-state hyperfine manifold),
$\Gamma_{3}=2\pi\times 5.75$ MHz, and $\Gamma_{4}=2\pi\times6.07$ MHz~\cite{Steck}.
% In the analytical derivations below, we measure length in units of the lattice constant and set ${\bf a}=1$.

%% fig3 %%%%%%%%%%%%%%%%%%%%%%%%%%%%%%%%%%%%%%%%%%%%%%%%%%%%%%%%%%%%%
\begin{figure*}[ht]
  \includegraphics[width=0.9\linewidth]{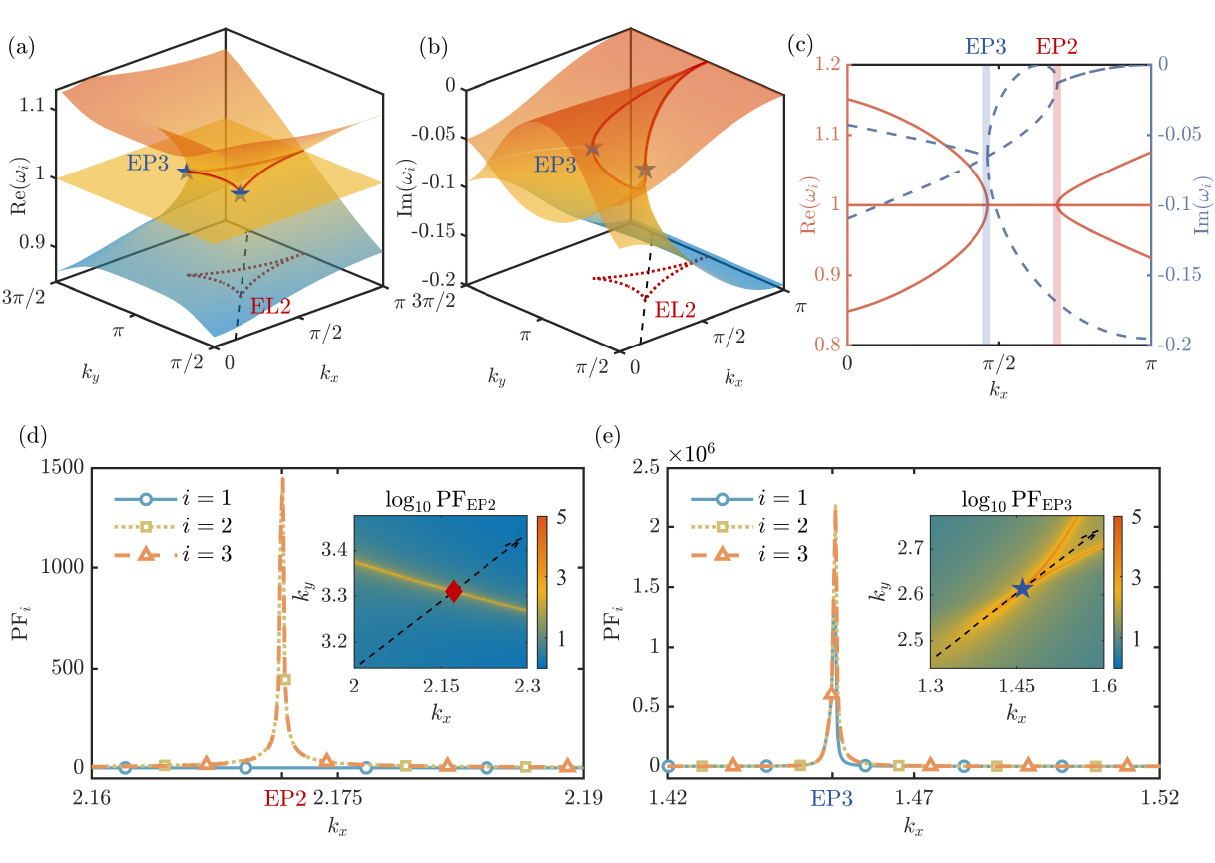}
  \caption{
(a) Real and (b) imaginary parts of the three eigenvalues $\omega_i$ as
functions of the wave vector $(k_x,k_y)$. Blue stars mark the EP3 locations,
and red lines indicate the EL2s and their
projections onto the bottom plane.
(c) 
% Cross-sectional eigenvalue profiles along the cut marked by the black dashed line in (a), 
Eigenvalue profiles along the black dashed cut in (a) and (b),
$k_y=k_x+1.15366$. Red solid and blue dashed
lines show the real and imaginary parts, respectively. The blue and red
shaded vertical bands mark the EP3 and EP2 at 
$k_x=1.45302$
and $2.16091$,
respectively.
(d) Petermann factors $\mathrm{PF}_i$ as
functions of $k_x$, with $k_y=3.31457$; the PFs of the two coalescing modes diverge.
(e) Same as in (d), but with $k_y=2.60668$, where all three PFs diverge. The insets of (d) and (e) show the logarithmic PF distributions in the $(k_x,k_y)$ plane near the EP2 and EP3. The red diamond and blue star mark the EP2 and EP3, respectively. The black dashed lines are the same as those in (a) and (b). Other parameters $\gamma=0.195$ and $\kappa=0.071$.
}
  \label{fig3}
  \end{figure*}
%% fig3 %%%%%%%%%%%%%%%%%%%%%%%%%%%%%%%%%%%%%%%%%%%%%%%%%%%%%%%%%%%%%

\section{Exceptional points and Petermann
factor}
\label{sec:complex_band}

\subsection{Exceptional points and exceptional lines}
\label{subsec:exceptional_lines}

We now analyze the spectrum of the non-Hermitian Lieb lattice. Since $\mathcal{H}$ is a $3\times3$ matrix, its band structure is determined by the roots of the characteristic polynomial
\begin{align}
P(\omega) = \det\!\left[\mathcal{H} - \omega\,\mathbb{I}\right]
= f_3\omega^3 + f_2\omega^2 + f_1\omega + f_0,\notag
% \label{eq:cubic_polynomial}
\end{align}
where $\mathbb{I}$ is the identity matrix, and the coefficients are
$f_3=-1$,
$f_2=3n_0-i\gamma$,
$f_1=-3n_0^2+2in_0\gamma+4\kappa^2
+2\kappa^2[\cos k_x+\cos k_y]$, and
$f_0=n_0^2(n_0-i\gamma)
-2\kappa^2(2n_0-i\gamma)
-2\kappa^2n_0\cos k_x
-2\kappa^2(n_0-i\gamma)\cos k_y$.
% The $x$--$y$ asymmetry of $f_0$ reflects the fact that the lossy $C$ sites
% couple to the lattice only through $g_x$.

An EP3, at which three eigenvalues and their eigenvectors coalesce, corresponds to a triple root of the characteristic polynomial and is identified by the conditions
\begin{equation}
P(\omega)=0,
\quad
P'(\omega)=0,
\quad
P''(\omega)=0.\notag
% \label{eq:EP3_triple_root}
\end{equation}
Since $P''(\omega)=6f_3\omega+2f_2$ is independent of $\mathbf{k}$, the last
condition fixes the coalesced eigenvalue at the spectral centroid,
\begin{equation}
\omega_{\mathrm{EP3}} = -\frac{f_2}{3f_3}
= n_0 - \frac{i\gamma}{3}.\notag
% \label{eq:EP3_eigenvalue}
\end{equation}
The remaining two conditions determine the momenta. The condition $P'(\omega_{\mathrm{EP3}})=0$ requires $g_x^2+g_y^2=\gamma^2/3$, while $P(\omega_{\mathrm{EP3}})=0$ gives $g_y^2=\gamma^2/27$ and  $g_x^2=8\gamma^2/27$.
% Using $g_{x,y}^2=2\kappa^2[1+\cos k_{x,y}]$,
Then we obtain
\begin{equation}
k_x
=
\pm\arccos\!\left[\frac{4\gamma^2-27\kappa^2}{27\kappa^2}\right],
~
k_y
=
\pm\arccos\!\left[\frac{\gamma^2-54\kappa^2}{54\kappa^2}\right],
\label{eq:EP3_momentum}
\end{equation}
in the first Brillouin zone (the corresponding $k_{x,\,y}\in [-\pi, \pi]$ with $a=1$).  
% Note that we take $a = 1$ throughout this paper.
The four sign combinations $(\pm k_x,\pm k_y)$ give four symmetry-related EP3 solutions in the first Brillouin zone.
Real solutions of Eq.~\eqref{eq:EP3_momentum} require $\gamma^2\leq27\kappa^2/2$.
% ; the constraint from $k_y$, $\gamma^2\leq108\kappa^2$, is less restrictive. 
For real $k_x$ and $k_y$, both inverse-cosine arguments in Eq.~\eqref{eq:EP3_momentum} must lie in $[-1,1]$, which requires $0<|\gamma|\leq\sqrt{27/2}\,\kappa$. For nonzero $\gamma$ in this range, the eigenspace associated with $\omega_{\mathrm{EP3}}$ is one-dimensional, confirming that all three eigenvectors coalesce and that the degeneracy is an EP3~\cite{demange2012signatures}. As $\gamma\to0$, the EP3s approach the Hermitian triple degeneracy at $\mathbf{k}=(\pi,\pi)$, where the flat band touches the two dispersive bands.

We also calculate the band structure numerically. Figures~\ref{fig3}(a) and (b) show the real and imaginary parts of the eigenvalues of $\mathcal{H}$ on the  plane $(k_x,k_y)$. In Fig.~\ref{fig3}(a), the real spectrum comprises a flat band and two dispersive bands. The panels show the EP3 locations (blue stars) and three  EL2s (red lines). 
The corresponding imaginary spectrum is shown in Fig.~\ref{fig3}(b).
The cross section in Fig.~\ref{fig3}(c) resolves the eigenvalues along the cut marked by the black dashed curve, $k_y=\,k_x+1.15366$; the real and imaginary parts are plotted as red solid and blue dashed lines, respectively. At $k_x=1.45302$, three eigenvalue branches coalesce, indicating EP3. EP2 occurs near $k_x=2.16091$, where two branches coalesce.

\subsection{Petermann factors}
\label{subsec:EP3_nonorthogonality}

Having established the conditions for EP2 and EP3, we now characterize the associated mode nonorthogonality. For the non-Hermitian Hamiltonian $\mathcal{H}$, the right and left eigenvectors satisfy $\mathcal{H}|\Psi_n^R\rangle=\omega_n|\Psi_n^R\rangle$ and $\mathcal{H}^\dagger|\Psi_n^L\rangle=\omega_n^*|\Psi_n^L\rangle$, respectively. 
% 
% For distinct eigenvalues, the two sets of eigenvectors obey the biorthogonality relation $\langle\Psi_n^L|\Psi_m^R\rangle=0$ for $n\ne m$. 
% Each nondegenerate pair can be normalized such that $\langle\Psi_n^L|\Psi_n^R\rangle=1$.
At an EP, 
% the coalescing right eigenvectors—and likewise the left eigenvectors—become linearly dependent.
the left--right self-overlap of each coalescing mode then vanishes, $\langle\Psi_n^L|\Psi_n^R\rangle\to0$. The resulting enhancement of mode nonorthogonality is quantified by the normalization-independent Petermann factor (PF)~\cite{NH30,ADD32_Lee2009}:
\begin{equation}
\mathrm{PF}_i
=
\frac{
\langle\Psi_i^L|\Psi_i^L\rangle
\langle\Psi_i^R|\Psi_i^R\rangle
}{
\left|
\langle\Psi_i^L|\Psi_i^R\rangle
\right|^2
}.
\label{eq:Petermann_general}
\end{equation}

The PF is invariant under independent rescalings of the left and right eigenvectors. The Cauchy--Schwarz inequality gives $\mathrm{PF}_i\geq1$, with equality when the left and right eigenvectors are proportional under the inner product used in Eq.~\eqref{eq:Petermann_general}. Thus, the PF quantifies modal nonorthogonality and diverges as the left--right overlap vanishes at an EP.

Near an EP2, the PFs of the two coalescing modes (orange  dashed and yellow dotted lines) diverge, whereas that of the remaining nondegenerate mode (blue solid line) remains finite; see Fig.~\ref{fig3}(d). At the EP3, all three PFs diverge, see Fig.~\ref{fig3}(e). These behaviors reflect the respective two- and three-mode coalescence. 
The insets of Fig.~\ref{fig3}(d) and (e) show the logarithmic PF distributions in the $(k_x,k_y)$ plane near the EP2 and EP3. The red diamond and blue star mark the EP2 and EP3, respectively. 
% The black dashed lines are the same as those in (a) and (b). 
% Although PF divergence signals strong modal nonorthogonality, it does not by itself determine the spontaneous-emission rate. The latter must be evaluated from the Green's function and the associated LDOS, as discussed in Sec.~\ref{sec:LDOS_enhancement}.

\section{Enhanced local density of states}
\label{sec:LDOS_enhancement}

Non-orthogonality of the modes in open systems can lead to significantly enhanced spontaneous-emission rates~\cite{petermann1979calculated}. This effect becomes most pronounced near EPs~\cite{ADD30_Berry2003}, where the modes become self-orthogonal.
The divergent PF discussed above reveal the
pronounced nonorthogonality of the eigenmodes near an EP. To determine how such mode coalescence modifies spontaneous
emission, we now examine the local density of states
(LDOS)~\cite{Lin2016EnhancedEmissionEP3,Pick2017GeneralTheory}.
In general, an emitter's spontaneous-emission rate is determined by the electromagnetic modes available at its position, as quantified by the LDOS~\cite{NH61,NH66,ADD34_Kristensen2020}. Because the LDOS is proportional to the negative imaginary part of the corresponding local Green's function element, we construct the Green's function representation for the non-Hermitian Lieb lattice.

In a non-Hermitian formulation, dissipation can be incorporated through effective complex on-site parameters. The Green's function describes the system's response to a localized point source. 
%%
% In continuous space, it satisfies $[H(\mathbf{r})-\omega\mathbb{I}]G(\mathbf{r},\mathbf{r}';\omega)=-\delta(\mathbf{r}-\mathbf{r}')$, where $H(\mathbf{r})$ is the non-Hermitian operator, $\omega$ is the excitation frequency, and $\delta(\mathbf{r}-\mathbf{r}')$ describes a point source at $\mathbf{r}'$.
%%
In the tight-binding representation, 
% we use localized lattice states as the basis. The equation then becomes 
it satisfies $(\mathcal{H}-\omega\,\mathbb{I})G(\omega)=-\mathbb{I}$, so the Green's function is $G(\omega)=(\omega\,\mathbb{I}-\mathcal{H})^{-1}$. 
% Here $\mathbb{I}$ is identity matrix.
%%
% The resonant modes satisfy the non-Hermitian eigenvalue equation $\mathcal{H}|\Psi_n^{R}\rangle=\omega_n|\Psi_n^{R}\rangle$, where $\omega_n$ and $|\Psi_n^{R}\rangle$ are the $n$th complex eigenvalue and right eigenvector, respectively. The corresponding left eigenvector satisfies $\langle\Psi_n^{L}|\mathcal{H}=\omega_n\langle\Psi_n^{L}|$.
%%
Away from an EP, the Hamiltonian is diagonalizable, and its left and right eigenvectors form a complete biorthogonal basis. The Green's function then has the modal expansion~\cite{ADD34_Kristensen2020,ADD37_Sauvan2022}
\begin{equation}
G(\omega)
=
\sum_n
\frac{
|\Psi_n^{R}\rangle
\langle\Psi_n^{L}|
}{
(\omega-\omega_n)
\langle\Psi_n^{L}|\Psi_n^{R}\rangle
}.
\label{eq:Green_modal_general}
\end{equation}
% $|\Psi_n^{R}\rangle $ and $\langle\Psi_n^{L}|$ are the right and left eigenvectors.

For the $j$th local site channel, the LDOS is proportional to the
negative imaginary part of the corresponding diagonal element of
the Green's function~\cite{Pick2017GeneralTheory},
\begin{equation}
\mathrm{LDOS}_{j}(\omega)
\propto
-\operatorname{Im}
\left[
G_{jj}(\omega)
\right].
\label{eq:LDOS_local_definition}
\end{equation} 
The tight-binding Green's function used here is expressed in a localized lattice channel basis and does not explicitly resolve the vector polarization of an arbitrary physical dipole. Thus, $G_{jj}$ gives the local spectral response of channel $j$; a quantitative prediction of a specific emitter's decay rate requires mapping its position and dipole orientation onto the corresponding lattice channel coupling.

Equation~\eqref{eq:Green_modal_general} applies when the Hamiltonian is diagonalizable. At an EP, the ordinary eigenvectors no longer form a complete basis, and the conventional modal expansion breaks down~\cite{NH64,NH70}. The Green's function must instead be reconstructed using the Jordan chain associated with the coalesced modes~\cite{Ferrier2022UnveilingEmission}. We treat EP2 and EP3 separately below because their Jordan chains and their pole structures differ.

\subsection{Second-order exceptional point}

At an EP2, two eigenmodes coalesce while the third remains nondegenerate. Because the ordinary eigenvectors no longer form a complete basis, an associated vector is needed to construct the second-order Jordan chain. 
The construction and normalization
of the left and right Jordan chains are detailed in
Appendix~\ref{app:Jordan_construction}.
% Appendix~\ref{app:Jordan_construction} details the construction and normalization of the left and right chains.

Using the normalized Jordan chains, the contribution of the
coalescing modes to the Green's function is
\begin{equation}
\begin{aligned}
G_{\mathrm{EP2}}(\omega)
={}&
\frac{|\Psi_0^R\rangle\langle\Psi_0^L|}
     {(\omega-\omega_{\mathrm{EP2}})^2}
% \\
% &
+
\frac{
|\Psi_0^R\rangle\langle J_1^L|
+
|J_1^R\rangle\langle\Psi_0^L|
}{
\omega-\omega_{\mathrm{EP2}}
}.
\end{aligned}
\label{eq:Green_EP2_Jordan}
\end{equation}
Here $\omega_{\rm EP2}$ is the second-order coalesce eigenvalue,
$|J_1^R\rangle$ and $\langle J_1^L|$ are Jordan vectors.
This expression contains first- and second-order poles. The second-order pole distinguishes the EP2 response from that of an isolated nondegenerate mode, which contributes only a first-order pole.

% For this three-band model, the complete Green's function also includes the remaining nondegenerate mode and can be written as
% $G_{\mathrm{EP2}}(\omega)
% =G_{\mathrm{EP2}}^{(\mathrm{J})}(\omega)
% +G_{\mathrm{nond}}(\omega)$, 
% {\color{red}
% where
% $G_{\mathrm{nond}}(\omega)
% =|\Psi_r^R\rangle\langle\Psi_r^L|
% \big/
% [(\omega-\omega_r)
% \langle\Psi_r^L|\Psi_r^R\rangle]$.
% When this contribution is small and varies slowly across the resonance range of interest, the local spectral response is dominated by $G_{\mathrm{EP2}}^{(\mathrm{J})}(\omega)$~\cite{}.
% }

\begin{figure}[t]
  \includegraphics[width=\linewidth]{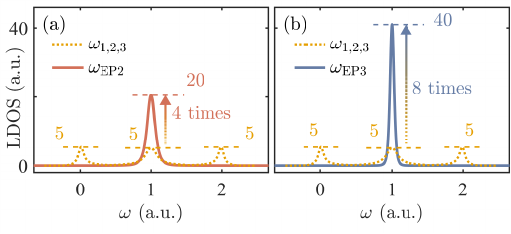}\caption{Local density of states (LDOS) spectra at and away from EPs.
(a) 
% LDOS near an EP2. 
Away from the EP, the three nondegenerate modes produce
spectrally separated resonances $\omega_{1,2,3}$ (yellow dashed line). 
% Here 
% $\gamma=0.195$, $\kappa=1.95$, and 
% $k_x=k_y=\pi/2$.
 % each with a peak value of $0.1$. 
At the EP2, two modes coalesce at $\omega=1$,
producing the resonance $\omega_{\mathrm{EP2}}$ (red solid line), whose peak
value is $20$, four times the peak of a single nondegenerate resonance with 
% $\gamma=0.195$, $\kappa=0.071$, 
$k_x=2.16091$ and $k_y=3.31457$.
(b) 
% LDOS near an EP3. 
Away from the EP, the three nondegenerate modes produce
three separated resonances same as (a). 
% $\gamma=0.195$, $\kappa=1.95$, 
At the EP3, all three modes coalesce at $\omega=1$,
producing the resonance $\omega_{\mathrm{EP3}}$ (blue solid line), whose peak
value is $40$, eight times the peak of a single nondegenerate resonance with $k_x=1.45302$ and $k_y=2.60668$.
% Breaks in the horizontal axes omit the intermediate frequency intervals between
% % the central and side resonances. Here $\gamma=0.195$, $\kappa=0.071$, 
% $k_x=1.45302$, and $k_y=2.60668$.
Other parameters $\gamma=0.195$, $\kappa=0.071$ for at the EP, $\kappa=0.5$ and $k_x=k_y=\pi/2$ for away from EP.
% \textcolor{red}{EP3 \gamma=0.195, \kappa=0.071, k_x=1.45303, k_y=2.60665;
% Far EP3 \gamma=0.195, \kappa=1.95, k_x=\pi/4, k_y=\pi/4;
% EP2  \gamma=0.195, \kappa=0.071, k_x=2.16091, k_y=3.31457;
% Far  EP2  \gamma=0.195, \kappa=1.95, k_x=\pi/2, k_y=\pi/2.
% }
% The total spectrally integrated intensity of
% the separated off-EP resonances equals that of the corresponding coalesced
% resonance. Coalescence concentrates the spectral response near the central
% frequency, producing a higher and narrower peak. These values apply to the
% local channels and model parameters considered here and are not universal
% spontaneous-emission enhancement factors.
}
  \label{fig4}
\end{figure}

For the second local channel considered in Fig.~\ref{fig4}(a), the LDOS is proportional to the negative imaginary part of the second diagonal element of the Green's function, i.e.,
% {\color{red}
% Under the resonance approximation~\cite{}, the selected configuration gives}
\begin{equation}
\begin{aligned}
\mathrm{LDOS}_{\mathrm{EP2}}(\omega)
&\propto
-\operatorname{Im}
\left[
G_{\mathrm{EP2}}(\omega)_{22}
\right]
\\
&\simeq
-\operatorname{Im}
\left[
\frac{i\gamma/2}
     {(\omega-\omega_{\mathrm{EP2}})^2}
+
\frac{1}
     {\omega-\omega_{\mathrm{EP2}}}
\right].
\end{aligned}
\label{eq:LDOS_EP2}
\end{equation}
The first- and second-order pole contributions jointly determine the local resonance lineshape and produce the enhanced peak shown in Fig.~\ref{fig4}(a). 
% Because $\omega_{\mathrm{EP2}}$ has a nonzero imaginary part, the poles lie off the real-frequency axis, and the local spectral response remains finite despite eigenmode coalescence.
Away from the EP, we find the three nondegenerate modes produce
spectrally separated resonances $\omega_{1,2,3}$ (yellow dashed lines), each with a peak value of $5$. 
At the EP2, two modes coalesce at $\omega=1$,
producing the resonance $\omega_{\mathrm{EP2}}$ (red solid line), whose peak
value is $20$, four times the peak of a single nondegenerate resonance. 
Hence   LDOS can achieve a 4 times enhancement in EP2 cases.

%% fig4 %%%%%%%%%%%%%%%%%%%%%%%%%%%%%%%%%%%%%%%%%%%%%%%%%%%%%%%%%%%%%
%% fig4 %%%%%%%%%%%%%%%%%%%%%%%%%%%%%%%%%%%%%%%%%%%%%%%%%%%%%%%%%%%%%

% We next apply the same Jordan-chain approach to the EP3 case.

\subsection{ Third-order exceptional point}

We now apply the same Jordan-chain approach to the EP3 case.
At an EP3, all three eigenmodes coalesce, and two associated vectors are needed to form a complete third-order Jordan chain. 
The details are shown in
Appendix~\ref{app:Jordan_construction}.
Using the normalized Jordan chains, the Green's function can be
rewritten as
\begin{equation}
\begin{aligned}
G_{\mathrm{EP3}}(\omega)
={}&
\frac{|\Psi_0^R\rangle\langle\Psi_0^L|}
     {\Delta\omega^3}+
\frac{
|\Psi_0^R\rangle\langle J_1^L|
+
|J_1^R\rangle\langle\Psi_0^L|
}{
\Delta\omega^2
}
\\
&+
\frac{
|\Psi_0^R\rangle\langle J_2^L|
+
|J_1^R\rangle\langle J_1^L|
+
|J_2^R\rangle\langle\Psi_0^L|
}{
\Delta\omega
},
\end{aligned}
\label{eq:Green_EP3_complete}
\end{equation}
where $\Delta\omega=\omega-\omega_{\mathrm{EP3}}$, with $\omega_{\rm EP3}$ is the third-order coalesce eigenvalue.
Unlike the EP2 case, all three modes belong to this Jordan chain,
% so no additional contribution from a nondegenerate mode is needed.
so Eq.~\eqref{eq:Green_EP3_complete} gives the
complete Green's function of the three-band Hamiltonian at the
EP3.

For the third local channel considered in Fig.~\ref{fig4}(b),
the LDOS is determined by 
% the negative imaginary part of the
% third diagonal element of the Green's function. Substituting
% the EP3 conditions into the Hamiltonian yields
\begin{equation}
\begin{aligned}
\mathrm{LDOS}_{\mathrm{EP3}}(\omega)
&\propto
-\operatorname{Im}
\left[
G_{\mathrm{EP3}}(\omega)_{11}
\right]
\\
&\simeq
-\operatorname{Im}
\left[
\frac{1}{\Delta\omega}
-
\frac{2i\gamma}{3\Delta\omega^2}
-
\frac{4\gamma^2}{27\Delta\omega^3}
\right].
\end{aligned}
\label{eq:LDOS_EP3}
\end{equation}
Compared with the EP2 response, the additional third-order pole further modifies the frequency dependence of the local Green's function. The first-, second-, and third-order terms jointly determine the LDOS lineshape, so the peak height cannot be inferred from the pole order alone. This spectrum underlies the local response shown in Fig.~\ref{fig4}(b).
 Away from the EP, the three nondegenerate modes (yellow dashed line) produce
three separated resonances. At the EP3, all three modes coalesce at $\omega=1$,
producing the resonance $\omega_{\mathrm{EP3}}$ (blue solid line), whose peak
value is $40$, eight times the peak of a single nondegenerate resonance. Hence   LDOS can achieve a 8 times enhancement in EP3 cases.

\begin{figure}[ht]
  \includegraphics[width=1\linewidth]
  {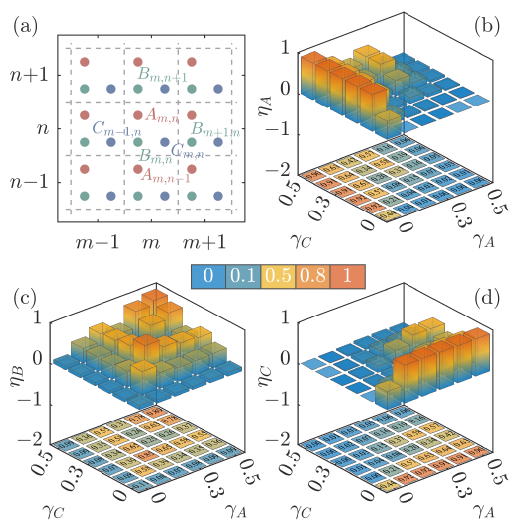}
  \caption{
  (a) Lieb lattice. Each unit cell $(m,n)$ contains three sites,
  $A_{m,n}$, $B_{m,n}$, and $C_{m,n}$, shown in red, green, and blue, respectively.
  (b)--(d) Relative field-intensity distribution of lattice sites $A$ ($\eta_A$), $B$ ($\eta_B$), and $C$ ($\eta_C$).  
  % (b) Relative field-intensity distribution $\eta_A$. (c) Relative field-intensity distribution $\eta_B$. (d) Relative field-intensity distribution $\eta_C$. 
  % $\eta_B(Z)$, and $\eta_C(Z)$ on the $A$, $B$, and $C$ sites,
  % respectively, at the propagation distance $Z=20$, plotted as functions
  % of the gain--loss parameters $\gamma_A,\gamma_C\in[-4,4]$, with
  % $\gamma_B=0$. The bar heights show the corresponding fractions, and the
  % color maps projected onto the bottom planes display their distributions
  % in parameter space. The complementary regions illustrate how the
  % sublattice-dependent gain--loss imbalance controls the relative local
  % intensity distribution among the three sites of the central unit cell. 
  The color maps projected onto the bottom planes display their distributions in parameter space. Here $z=20$, $\gamma_B=0.195$, and $\kappa=0.071$. 
  } 
  \label{fig5}
\end{figure}
%% fig5 %%%%%%%%%%%%%%%%%%%%%%%%%%%%%%%%%%%%%%%%%%%%%%%%%%%%%%%%%%%%%

\section{Localization }
\label{sec:propagation_dynamics}

% Propagation characteristics of the non-Hermitian Lieb photonic lattice are obtained by solving the Maxwell equation.
% Under the slowly varying envelope approximation, the signal field obeys the equation~\cite{Qin2019ControllablePT}:
% \begin{equation}\label{eq:Maxwell}
% i\left(\frac{\partial}{\partial z}+\frac{1}{c}\frac{\partial}{\partial t}\right)\Omega_s+\frac{c}{2\omega_s}\left(\frac{\partial^2}{\partial x^2}+\frac{\partial^2}{\partial y^2}\right)\Omega_s+\frac{\omega_s}{2c}\chi_s\Omega_s=0,
% \end{equation}
%  We consider the steady-state regime, in which the time derivatives in Eq.~\eqref{eq:Maxwell} can be neglected. 
% The preceding analysis shows that the atomic medium's complex  susceptibility provides tunable complex on-site parameters for the Lieb lattice. Its real part primarily controls the local dispersive refractive-index modulation, whereas its imaginary part introduces gain or loss on selected sublattices. Having examined the resulting band structure, EPs, PFs, and LDOS, we now use the nearest-neighbor tight-binding model to investigate how these on-site parameters affect real-space propagation and redistribute the field intensity among the sites of the central unit cell~\cite{NH47}.

We also study wave dynamics in different loss scenarios by using the tight-binding model. Tight-binding model is quite accurate for describing the two-dimensional Lieb lattice~\cite{Leykam2012}, which is described as
% While it does have simplifications on ignoring the fine structures inside each lattice site, it grabs the essential properties of the whole lattice, i.e., the average refractive index together with the representative loss/gain of each site and the coupling between neighboring sites.
% The wave dynamics in the non-Hermitian Lieb lattice with the tight-binding model is described as:
\begin{subequations}
\begin{align}
i\frac{\partial A_{m,n}}{\partial z}
={}&
-i\gamma_A A_{m,n}
+
\kappa
\left(
B_{m,n}+B_{m,n+1}
\right),
\label{eq:propagation_A}
\\
i\frac{\partial B_{m,n}}{\partial z}
={}&
-i\gamma_B B_{m,n}
+
\kappa
\left(
A_{m,n}+A_{m,n-1}
\right)
\nonumber\\
&+
\kappa
\left(
C_{m,n}+C_{m-1,n}
\right),
\label{eq:propagation_B}
\\
i\frac{\partial C_{m,n}}{\partial z}
={}&
-i\gamma_C C_{m,n}
+
\kappa
\left(
B_{m,n}+B_{m+1,n}
\right).
\label{eq:propagation_C}
\end{align}
\end{subequations}
Here, $A_{m,n}$, $B_{m,n}$, $C_{m,n}$ are the fields at sites
$m$ rows and $n$ columns as shown in Fig.~\ref{fig5}(a).
$\kappa$ is the nearest-neighbor coupling strength, and $\gamma_A$, $\gamma_B$, and $\gamma_C$ are
loss for the three sublattices. 
% $k_x=k_y=0$.
% With the sign convention in Eqs.~\eqref{eq:propagation_A}--\eqref{eq:propagation_C}, $\gamma_j>0$, $\gamma_j<0$, and $\gamma_j=0$ correspond to loss, gain, and a neutral response, respectively, on sublattice $j\in\{A,\,B,\,C\}$.

%% fig6%%%%%%%%%%%%%%%%%%%%%%%%%%%%%%%%%%%%%%%%%%%%%%%%%%%%%%%%%%%%%
\begin{figure*}[ht]
  \includegraphics[width=1\linewidth]
  {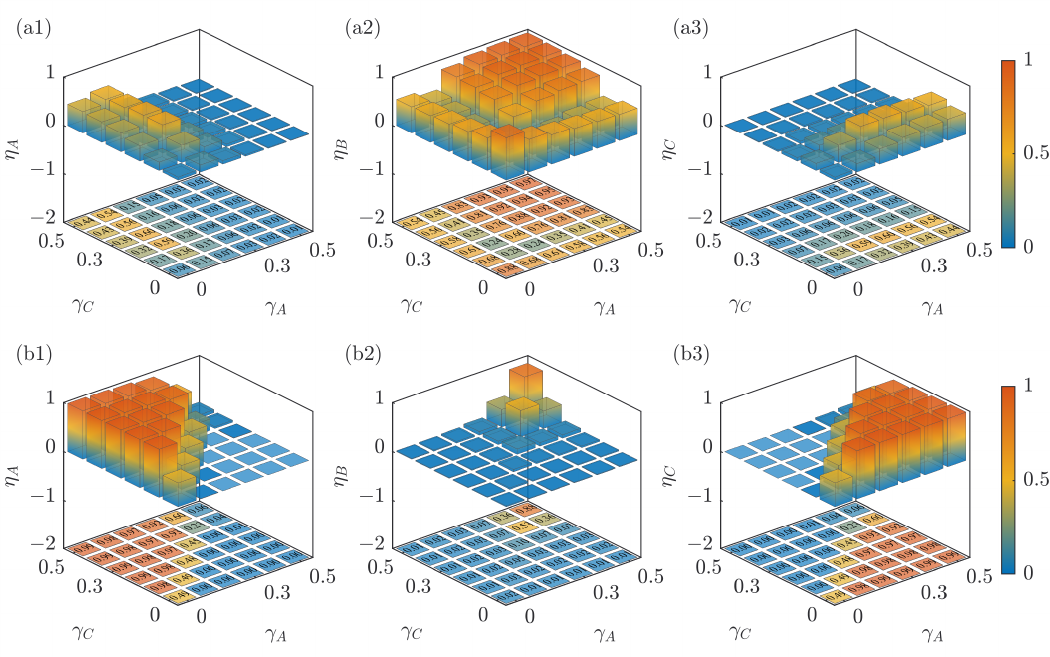}
  \caption{
  The same as Fig.~\ref{fig5}(b)-(d), relative field-intensity distribution of lattice sites $A$, $B$, and $C$.    (a1)-(a3) A neutral response at $B$ site, with $\gamma_b=0$. (b1)-(b3) A larger loss  at $B$ site, with $\gamma_b=0.5$.
  } 
  \label{figS1}
\end{figure*}
%% figS1 %%%%%%%%%%%%%%%%%%%%%%%%%%%%%%%%%%%%%%%%%%%%%%%%%%%%%%%%%%%%%

For the numerical calculations, we use a finite Lieb lattice of $25\times25$ unit cells with open boundary conditions. We set the nearest-neighbor coupling strength to \(\kappa=0.071\) and keep the B sublattice weakly lossy, i.e. \(\gamma_B=0.195\), while varying \(\gamma_A\) and \(\gamma_C\) independently over \([0,\,0.5]\). We integrate the propagation equations using a fourth-order Runge--Kutta method~\cite{yang2010nonlinear}.
% .
At $z=0$, we apply an equal-amplitude, in-phase coherent excitation to the three sites in the central unit cell, i.e. $m,\,n=0$, so that $A_{0,0}(z=0)=B_{0,0}(z=0)=C_{0,0}(z=0)=1$, while all other sites are unexcited. This initial condition lets us track the evolution of the three local sublattice amplitudes under the combined action of intersite coupling and non-Hermitian on-site terms.

To characterize the relative field-intensity distribution within the central unit cell at a fixed propagation distance, we define
 % the normalized local intensities
% \begin{equation*}
% \begin{aligned}
$\eta_A(z)
=
{|A_{0,0}(z)|^2}/
{\mathcal{I}(z)}$,
$\eta_B(z)
=
{|B_{0,0}(z)|^2}/
{\mathcal{I}(z)}$, 
and
$\eta_C(z)
=
{|C_{0,0}(z)|^2}/
{\mathcal{I}(z)}$.
% \end{aligned}
% \label{eq:normalized_local_intensities}
% \end{equation*}
Here $z$ is the propagation distance, 
and $\mathcal{I}(z)=|A_{0,0}(z)|^2+|B_{0,0}(z)|^2+|C_{0,0}(z)|^2$, so that $\eta_A(z)+\eta_B(z)+\eta_C(z)=1$. These quantities describe the relative intensity distribution among the three sites in the central unit cell.

Figure~\ref{fig5}(b)–(d) show the relative field intensities $\eta_A$, $\eta_B$, and $\eta_C$ within the central unit cell as functions of the loss parameters $\gamma_A$ and $\gamma_C$, with $\gamma_B=0.195$. 
The values in the projection table represent the corresponding relative field intensities.
As shown in Fig.~\ref{fig5}(b), $\eta_A$ is relatively high when $\gamma_A$ is small and $\gamma_C$ is large, indicating that lower local loss favors the relative retention of the field at the $A$ site. Figure~\ref{fig5}(d) shows a corresponding trend for the $C$ site: when $\gamma_C$ is small and $\gamma_A$ is large, the local intensity is concentrated mainly at the $C$ site.

As shown in Fig.~\ref{fig5}(c), $\eta_B$ increases significantly when both $\gamma_A$ and $\gamma_C$ are large. In this region, the $B$ sublattice has relatively lower loss, so the field on this sublattice experiences weaker attenuation than those on the other two sublattices. When the loss on the $A$ or $C$ sublattice is smaller, the corresponding site retains a larger intensity fraction, reducing $\eta_B$.

Similarly, tuning the loss at the B site also changes the localization pattern. For example, for $\gamma_b = 0$ [a neutral], the state localizes mainly at the B site as shown in Fig.~\ref{figS1}(a2). 
When the loss on the $A$ sublattice is smaller and $C$ is higher, the $A$ and $B$ sites retain a larger intensity fraction. When the loss on the $C$ sublattice is smaller and $A$ is higher, the $C$ and $B$ sites retain a larger intensity fraction.
For a larger loss at $B$, $\gamma_b = 0.5$, the localization moves to the $A$ site when the relative loss at $A$ is small [Fig.~\ref{figS1}(b1)], or to the $C$ site when the relative loss at $C$ is small [Fig.~\ref{figS1}(b3)].

These parameter-dependent distributions arise from the combined effects of local amplification or attenuation, nearest-neighbor coupling, and propagation into adjacent unit cells. The loss terms govern the local attenuation at each site, while the coupling terms redistribute the field among the three local components. The boundaries between regions of dominant response therefore depend not only on the relative magnitudes of \(\gamma_A\) and \(\gamma_C\), but also on the coupling strength and the propagation distance.
% }

%% 五、总结 %%%%%%%%%%%%%%%%%%%%%%%%%%%%%%%%%%%%%%%%%%%%%%%%%%%%%%%%%%%%%

%% 五、总结 %%%%%%%%%%%%%%%%%%%%%%%%%%%%%%%%%%%%%%%%%%%%%%%%%%%%%%%%%%%%%
\section{Conclusion}\label{conclusion}

% In this work, we have proposed and analyzed a physical scheme for realizing a non-Hermitian Lieb lattice in an optically pumped four-level $N$-type cold atom medium. By exploiting the tunable complex susceptibility of the atomic response, sublattice dependent dispersive and dissipative on-site parameters can be engineered in a three-band tight-binding framework, providing a natural platform for higher-order non-Hermitian degeneracies. Within this model, we identified second-order ELs and EP3s, and characterized the associated eigenvector coalescence through Petermann factors and normalized Jordan chains. A general theory of spontaneous emission at exceptional points was developed based on the local Green's function and the local density of states, showing that the spontaneous emission can be enhanced by factors of $4$ and $8$ at the EP2 and EP3, respectively. We further discussed localization in non-Hermitian lattices and showed that it can be realized at arbitrary lattice sites by suitably tuning the gain and loss of the lattice sites. Overall, this work demonstrates a controllable route to higher-order EP physics in a Lieb lattice setting and suggests practical applications in quantum sensing and imaging that require enhanced and controlled emission.

We have proposed a scheme for realizing a non-Hermitian Lieb lattice in an optically pumped four-level $N$-type cold atom medium. The tunable atomic susceptibility provides sublattice dependent dispersive and dissipative on-site terms in a three-band tight-binding model. We derived analytical conditions for an EP3 and identified EP3s and second-order ELs. The mode coalescence is characterized by the PF and by normalized Jordan chains, which also provide the Green's functions needed to evaluate the local spectral response. For the channels and parameters considered, the calculated LDOS peaks at the EP2 and EP3 are four and eight times, respectively, 
% the peak of a single nondegenerate resonance; these values are not universal spontaneous-emission enhancement factors. 
showing that the spontaneous-emission can be enhanced by factors of $4$ and $8$ in the passive system at the EP2 and EP3.
We also showed how tuning sublattice loss redistributes the relative intensity within a unit cell by tuning the relative loss. This work establishes a controllable platform for studying higher-order EPs and local optical responses in a non-Hermitian Lieb lattice.

\begin{acknowledgments}
L. Q., Y. Z., S. Z., and C. H. acknowledge the support of National Natural Science Foundation of China (Grant No. 12404377, No. 12304357, No. 62505079, and No. 12374303), the Program of China Scholarship Council under Grants No. 202508410217, and Shanghai Municipal Science and Technology Major Project (Grant No. 2019SHZDZX01).
 X. Z. acknowledges the support from the Key Scientific Research Project of Colleges and Universities in Henan Province (Grant No. 26B140007) and Natural Science Foundation of Henan Province (Grant No. 252300421995). Z. Z. acknowledges the support from the Key International Cooperation Project in Henan Province (Grant No. 261111521200). 
\end{acknowledgments}

\vspace{1mm}

\textit{Data availability.} The data that support the findings of this work are openly available \cite{code}.

\vspace{1mm}

\textit{AI usage statement.} The authors used \emph{OpenAI}
% 's \emph{Chatgpt solar} 
for grammar and spell-checking of the manuscript. All suggestions were reviewed and approved by the authors.

%% 六、附录 %%%%%%%%%%%%%%%%%%%%%%%%%%%%%%%%%%%%%%%%%%%%%%%%%%%%%%%%%%%%%

\appendix

%% 附录A %%%%%%%%%%%%%%%%%%%%%%%%%%%%%%%%%%%%%%%%%%%%%%%%%%%%%%%%%%%%%

% \onecolumngrid

\section{Optical Bloch equations}
\label{app:bloch_solution}

The optical Bloch equation describing the time evolution of the density-matrix elements $\rho_{\alpha\beta}$ reads~\cite{Qin2019ControllablePT}
\begin{subequations}
\begin{align}
&i\frac{\partial}{\partial t}\rho_{22}
-i\Gamma_{24}\rho_{44}
-i\Gamma_{23}\rho_{33}
+\Omega_c^*\rho_{32}
-\Omega_c\rho_{32}^*=0,
\label{eq:bloch_22}
\\
&i\frac{\partial}{\partial t}\rho_{33}
+i\Gamma_3\rho_{33}
+\Omega_s\rho_{31}^*
-\Omega_s^*\rho_{31}
+\Omega_c\rho_{32}^*
-\Omega_c^*\rho_{32}=0,
\label{eq:bloch_33}
\\
&i\frac{\partial}{\partial t}\rho_{44}
+i\Gamma_4\rho_{44}
+\Omega_p\rho_{41}^*
-\Omega_p^*\rho_{41}=0,
\label{eq:bloch_44}
\\
&\left(
i\frac{\partial}{\partial t}+d_{21}
\right)\rho_{21}
+\Omega_c^*\rho_{31}
-\Omega_s\rho_{32}^*
-\Omega_p\rho_{42}^*=0,
\label{eq:bloch_21}
\\
&\left(
i\frac{\partial}{\partial t}+d_{31}
\right)\rho_{31}
+\Omega_s\left(\rho_{11}-\rho_{33}\right)
+\Omega_c\rho_{21}
-\Omega_p\rho_{43}^*=0,
\label{eq:bloch_31}
\\
&\left(
i\frac{\partial}{\partial t}+d_{41}
\right)\rho_{41}
+\Omega_p\left(\rho_{11}-\rho_{44}\right)
-\Omega_s\rho_{43}=0,
\label{eq:bloch_41}
\\
&\left(
i\frac{\partial}{\partial t}+d_{32}
\right)\rho_{32}
+\Omega_c\left(\rho_{22}-\rho_{33}\right)
+\Omega_s\rho_{21}^*=0,
\label{eq:bloch_32}
\\
&\left(
i\frac{\partial}{\partial t}+d_{42}
\right)\rho_{42}
-\Omega_c\rho_{43}
+\Omega_p\rho_{21}^*=0,
\label{eq:bloch_42}
\\
&\left(
i\frac{\partial}{\partial t}+d_{43}
\right)\rho_{43}
+\Omega_p\rho_{31}^*
-\Omega_s^*\rho_{41}
-\Omega_c^*\rho_{42}=0.
\label{eq:bloch_43}
\end{align}
\end{subequations}
Here the superscript $*$ denotes complex conjugation, $d_{\alpha\beta}=\Delta_{\alpha}-\Delta_{\beta}+i\gamma_{\alpha\beta}$~$(\beta<\alpha)$, $\gamma_{\alpha\beta}=(\Gamma_{\alpha}+\Gamma_{\beta})/2+\gamma_{\alpha\beta}^{\mathrm{dep}}$, 
where $\Gamma_{\beta}=\sum_{\alpha<\beta}\Gamma_{\alpha\beta}$ is
spontaneous-emission rate and
 $\gamma_{\alpha\beta}^{\mathrm{dep}}$ is the additional dephasing rate. The populations are normalized according to $\sum_{\alpha=1}^{4}\rho_{\alpha\alpha}=1$~\cite{Boyd2008}.

\section{Solution of the Optical Bloch Equations}
\label{APP:solution}

We consider the stationary response of the atomic medium,
% . For sufficiently long laser pulses, transient dynamics can be neglected, 
so all time derivatives in Eqs.~\eqref{eq:bloch_22}--\eqref{eq:bloch_43} set to zero. Because the signal field is much weaker than the control and pump fields, we introduce a  perturbation parameter $\epsilon$ and take $\Omega_s=\epsilon\Omega_s^{(1)}$. 
Assuming the expansion $\rho_{\alpha\beta}=\rho_{\alpha\beta}^{(0)}+\epsilon\rho_{\alpha\beta}^{(1)}+\epsilon^2\rho_{\alpha\beta}^{(2)}+\cdots$, and substituting the expansion into the Bloch equation, we obtain a chain of linear but inhomogeneous equations, which can be solved order by order.

{\flushleft{\bf Zeroth order equations}}:
At the zeroth order, $\rho_{11}^{(0)}$, $\rho_{22}^{(0)}$, $\rho_{33}^{(0)}$, $\rho_{44}^{(0)}$, $\rho_{41}^{(0)}$, and $\rho_{32}^{(0)}$  can be solved by the equations
\begin{subequations}
\begin{align}
&\begin{pmatrix}
-i\Gamma_{24} & -i\Gamma_{23} & 0
& 0 & \Omega_c^* & 0 & -\Omega_c
\\
0 & i\Gamma_3 & 0
& 0 & -\Omega_c^* & 0 & \Omega_c
\\
i\Gamma_4 & 0 & 0
& -\Omega_p^* & 0 & \Omega_p & 0
\\
-2\Omega_p & -\Omega_p & -\Omega_p
& d_{41} & 0 & 0 & 0
\\
0 & -\Omega_c & \Omega_c
& 0 & d_{32} & 0 & 0
\\
2\Omega_p^* & \Omega_p^* & \Omega_p^*
& 0 & 0 & -d_{41}^* & 0
\\
 0 & \Omega_c^* & -\Omega_c^*
& 0 & 0 & 0 & -d_{32}^*
\end{pmatrix}
\begin{pmatrix}
\rho_{44}^{(0)}
\\
\rho_{33}^{(0)}
\\
\rho_{22}^{(0)}
\\
\rho_{41}^{(0)}
\\
\rho_{32}^{(0)}
\\
\rho_{41}^{*(0)}
\\
\rho_{32}^{*(0)}
\end{pmatrix}\notag\\
&\qquad=
\begin{pmatrix}
0
&
0
&
0
&
-\Omega_p
&
0
&
\Omega_p^*
&
0
\end{pmatrix}^{T},\\
\label{eq:zero_matrix}
&~\rho_{11}^{(0)}+\rho_{22}^{(0)}+\rho_{33}^{(0)}+\rho_{44}^{(0)}=1.
\end{align}
\end{subequations}
Other $\rho_{\alpha\beta}^{(0)}$ are zero.

{\flushleft{\bf First order equations}}:
% \noindent\textit{(ii) First order.}
At first order, the weak signal field induces the atomic coherences responsible for the linear optical response. The relevant density-matrix elements satisfy
\begin{align}
&\begin{pmatrix}
0 & \Omega_c^* & 0 & d_{21}
& 0 & 0 & -\Omega_p & 0
\\
0 & d_{31} & 0 & \Omega_c
& -\Omega_p & 0 & 0 & 0
\\
-\Omega_c & 0 & d_{42} & 0
& 0 & 0 & 0 & \Omega_p
\\
d_{43} & 0 & -\Omega_c^* & 0
& 0 & \Omega_p & 0 & 0
\\
0 & 0 & -\Omega_p^* & 0
& 0 & \Omega_c & 0 & d_{21}^*
\\
-\Omega_p^* & 0 & 0 & 0
& 0 & d_{31}^* & 0 & \Omega_c^*
\\
0 & 0 & 0 & \Omega_p^*
& -\Omega_c^* & 0 & d_{42}^* & 0
\\
0 & \Omega_p^* & 0 & 0
& d_{43}^* & 0 & -\Omega_c & 0
\end{pmatrix}
\begin{pmatrix}
\rho_{43}^{(1)}
\\
\rho_{31}^{(1)}
\\
\rho_{42}^{(1)}
\\
\rho_{21}^{(1)}
\\
\rho_{43}^{*(1)}
\\
\rho_{31}^{*(1)}
\\
\rho_{42}^{*(1)}
\\
\rho_{21}^{*(1)}
\end{pmatrix}\notag\\
&\qquad=
\begin{pmatrix}
\Omega_s\rho_{32}^{*(0)}
\\
\Omega_s
\left(\rho_{33}^{(0)}-\rho_{11}^{(0)}\right)
\\
0
\\
\Omega_s^*\rho_{41}^{(0)}
\\
\Omega_s^*\rho_{32}^{(0)}
\\
\Omega_s^*
\left(\rho_{33}^{(0)}-\rho_{11}^{(0)}\right)
\\
0
\\
\Omega_s\rho_{41}^{*(0)}
\end{pmatrix},
\label{eq:first_matrix}
\end{align}
and other $\rho_{\alpha\beta}^{(1)}$ are zero.

As we all known, with increasing expansion order, the higher-order density matrix elements become increasingly small. In this work, we mainly focus on linear non-Hermiticity, so we expand the density matrix elements only to first order.
% , i.e $\rho_{\alpha\beta}\approx\rho_{\alpha\beta}^{(0)}+\rho_{\alpha\beta}^{(1)}$.

\twocolumngrid

\section{Construction of Jordan chains}
\label{app:Jordan_construction}

% This appendix describes the construction and normalization of the left and right Jordan chains used in Sec.~\ref{sec:LDOS_enhancement}. The vectors $|\Psi_0^{R,L}\rangle$ denote the coalesced right and left eigenvectors, respectively, and $|J_j^{R,L}\rangle$ denote the associated vectors. The left eigenvectors and associated vectors are written in ket form and satisfy equations involving the adjoint Hamiltonian; their Hermitian conjugates give the corresponding bra vectors.

% The Green’s function can be expressed using the modal expansion formula as shown in Eq.~\eqref{eq:Green_modal_general}. 

At an EP, the Hamiltonian $\mathcal{H}$ does not have a complete basis of eigenvectors. Therefore, we must find an eigenvector  and an associated Jordan vector to diagonalize. For an EP2, eigenvector $|\Psi_0^R\rangle$ and an associated Jordan vector $|J_1^R\rangle$ satisfy the chain relations~\cite{mailybaev1999singularities,demange2012signatures}
\begin{subequations}
\label{eq:EP2_right_Jordan_chain}
\begin{align}
\mathcal{H}|\Psi_0^R\rangle
&=\omega_{\rm EP2}|\Psi_0^R\rangle,
\\
\mathcal{H}|J_1^R\rangle
&=\omega_{\rm EP2}|J_1^R\rangle+|\Psi_0^R\rangle,
\end{align}
\end{subequations}
where $\omega_{\rm EP2}$ is the degenerate eigenvalue.
Equivalent expressions are also available for the left eigenvector $|\Psi_0^L\rangle$ and Jordan vector $|J_1^L\rangle$.
In order to uniquely define the Jordan chain vectors, we need to specify self-orthogonality condition $\langle\Psi_0^{L}|\Psi_0^{R}\rangle=0$, 
and normalization condition
% , which we choose to be $\langle\Psi_0^{L}|\Psi_0^{R}\rangle=0$,
$\langle\Psi_0^{L}|J_1^{R}\rangle=1$, and $\langle J_1^L|J_1^R\rangle=0$.

% \subsection{Second-Order Exceptional Point}
% \label{app:Jordan_EP2}

% At an EP2, let $H_{\mathrm{EP2}}=H(\mathbf{k}_{\mathrm{EP2}})$ and $\omega_0=\omega_{\mathrm{EP2}}$. The right Jordan chain consists of the coalesced eigenvector $|\Psi_0^R\rangle$ and an associated vector $|J_1^R\rangle$, which satisfy
% \begin{subequations}
% \label{eq:EP2_right_Jordan_chain}
% \begin{align}
% H_{\mathrm{EP2}}|\Psi_0^R\rangle
% &=\omega_0|\Psi_0^R\rangle,
% \\
% H_{\mathrm{EP2}}|J_1^R\rangle
% &=\omega_0|J_1^R\rangle+|\Psi_0^R\rangle.
% \end{align}
% \end{subequations}

% The corresponding left Jordan chain comprises the left eigenvector $|\Psi_0^L\rangle$ and the associated vector $|J_1^L\rangle$, which satisfy
% $H_{\mathrm{EP2}}^\dagger|\Psi_0^L\rangle
% =\omega_0^*|\Psi_0^L\rangle$ and
% $H_{\mathrm{EP2}}^\dagger|J_1^L\rangle
% =\omega_0^*|J_1^L\rangle+|\Psi_0^L\rangle$.

% The chain relations do not uniquely determine the chain vectors. 
To satisfy the normalization conditions used in the Green's function expressions in the main text, we define transformed right-chain vectors:
\begin{subequations}
\label{eq:EP2_right_chain_transformation}
\begin{align}
|\widetilde{\Psi}_0^R\rangle
&=c_0|\Psi_0^R\rangle,
\\
|\widetilde{J}_1^R\rangle
&=c_0|J_1^R\rangle+c_1|\Psi_0^R\rangle.
\end{align}
\end{subequations}
Here, $c_0\neq0$ and $c_1$ is a complex coefficient.
% The left Jordan chain undergoes analogous transformations.
The new left and right chain vectors retain the original
Jordan chain relations, and the transformation coefficients
are chosen so that they also satisfy the normalization conditions.

% \begin{equation}
% \begin{pmatrix}
% \langle\widetilde{\Psi}_0^L|
%        \widetilde{\Psi}_0^R\rangle
% &
% \langle\widetilde{\Psi}_0^L|
%        \widetilde{J}_1^R\rangle
% \\
% \langle\widetilde{J}_1^L|
%        \widetilde{\Psi}_0^R\rangle
% &
% \langle\widetilde{J}_1^L|
%        \widetilde{J}_1^R\rangle
% \end{pmatrix}
% =
% \begin{pmatrix}
% 0&1\\
% 1&0
% \end{pmatrix}.
% \label{eq:appendix_EP2_normalization}
% \end{equation}

% This antidiagonal biorthogonal normalization makes the overlap between the coalesced left and right eigenvectors vanish, expressing the self-orthogonality of the coalesced mode. The two antidiagonal overlaps are unity, while the overlap between the associated left and right vectors is zero. The tildes are omitted from these normalized Jordan vectors in the main text.

% \subsection{Third-Order Exceptional Point}
% \label{app:Jordan_EP3}

At an EP3, the right Jordan chain comprises the coalesced eigenvector $|\Psi_0^R\rangle$ and two associated vectors, $|J_1^R\rangle$ and $|J_2^R\rangle$, which satisfy
\begin{subequations}
\label{eq:EP3_right_Jordan_chain}
\begin{align}
\mathcal{H}|\Psi_0^R\rangle
&=\omega_{\rm EP3}|\Psi_0^R\rangle,
\\
\mathcal{H}|J_1^R\rangle
&=\omega_{\rm EP3}|J_1^R\rangle+|\Psi_0^R\rangle,
\\
\mathcal{H}|J_2^R\rangle
&=\omega_{\rm EP3}|J_2^R\rangle+|J_1^R\rangle.
\end{align}
\end{subequations}
Similarly, we can write down the equation satisfied by the left eigenvector $|\Psi_0^L\rangle$ and two associated vectors, $|J_1^L\rangle$ and $|J_2^L\rangle$ eigenvector.
In order to uniquely define the Jordan chain vectors, we need satisfy orthogonality relations $\langle\Psi_0^{L}|\Psi_0^{R}\rangle=0$,  $\langle\Psi_0^{L}|J_1^{R}\rangle=0$, and $\langle\Psi_0^{L}|J_2^R\rangle=\langle J_1^L|J_1^R\rangle$, and 
additional normalization conditions
% , which we choose to be
% $\langle\Psi_0^{L}|\Psi_0^{R}\rangle=0$,
% $\langle\Psi_0^{L}|J_1^{R}\rangle=0$, 
$\langle\Psi_0^{L}|J_2^{R}\rangle=1$, 
% $\langle\Psi_0^{L}|J_2^R\rangle=\langle J_1^L|J_1^R\rangle$, 
and $\langle J_2^L|J_1^R\rangle=\langle J_2^L|J_2^R\rangle=0$~\cite{demange2012signatures}.

% The corresponding left Jordan chain comprises the left eigenvector $|\Psi_0^L\rangle$ and two associated vectors, $|J_1^L\rangle$ and $|J_2^L\rangle$, which satisfy
% $H_{\mathrm{EP3}}^\dagger|\Psi_0^L\rangle
% =\omega_0^*|\Psi_0^L\rangle$,
% $H_{\mathrm{EP3}}^\dagger|J_1^L\rangle
% =\omega_0^*|J_1^L\rangle+|\Psi_0^L\rangle$, and
% $H_{\mathrm{EP3}}^\dagger|J_2^L\rangle
% =\omega_0^*|J_2^L\rangle+|J_1^L\rangle$.

As in the EP3 case, to satisfy the normalization conditions
used in the Green's function expressions,
we introduce new right-chain vectors:
\begin{subequations}
\label{eq:EP3_right_chain_transformation}
\begin{align}
|\widetilde{\Psi}_0^R\rangle
&=c_0|\Psi_0^R\rangle,
\\
|\widetilde{J}_1^R\rangle
&=c_0|J_1^R\rangle+c_1|\Psi_0^R\rangle,
\\
|\widetilde{J}_2^R\rangle
&=c_0|J_2^R\rangle+c_1|J_1^R\rangle+c_2|\Psi_0^R\rangle.
\end{align}
\end{subequations}
Here, $c_0\neq0$, and $c_1$ and $c_2$ are complex
coefficients. 
% The left Jordan chain undergoes analogous
% transformations. 
The new left and right chain vectors retain
the original Jordan chain relations, and the transformation
coefficients are chosen so that they also satisfy the
normalization conditions.

\end{document}